\documentclass[iop,revtex4]{emulateapj}
\usepackage{color}

\newcommand{\ci}{\ion{C}{1}}
\newcommand{\ois}{\ion{O}{1}*}
\newcommand{\oiss}{\ion{O}{1}**}
\newcommand{\oi}{\ion{O}{1}}
\newcommand{\cet}{49~Cet}
\newcommand{\bp}{$\beta$~Pic}

\newcommand{\kms}{km~s$^{-1}$}
\newcommand{\tpz}{$^3\mathrm{P}_0$}
\newcommand{\tpo}{$^3\mathrm{P}_1$}
\newcommand{\tpt}{$^3\mathrm{P}_2$}
\newcommand{\cm}{cm$^{-2}$}

\shorttitle{Volatile Gases in the 49~Cet Disk}
\shortauthors{Roberge et al.}

\begin{document}

%% LaTeX will automatically break titles if they run longer than
%% one line. However, you may use \\ to force a line break if
%% you desire.

\title{Volatile-Rich Circumstellar Gas in the Unusual 49~Ceti Debris Disk}

%% Use \author, \affil, and the \and command to format
%% author and affiliation information.
%% Note that \email has replaced the old \authoremail command
%% from AASTeX v4.0. You can use \email to mark an email address
%% anywhere in the paper, not just in the front matter.
%% As in the title, use \\ to force line breaks.

\author{Aki Roberge\altaffilmark{1}, Barry Y.\ Welsh\altaffilmark{2,3}, Inga Kamp\altaffilmark{4}, Alycia J.\ Weinberger\altaffilmark{5}, \& Carol A.\ Grady\altaffilmark{2,1}}

%% Notice that each of these authors has alternate affiliations, which
%% are identified by the \altaffilmark after each name.  Specify alternate
%% affiliation information with \altaffiltext, with one command per each
%% affiliation.

\altaffiltext{1}{Exoplanets \& Stellar Astrophysics Laboratory, NASA Goddard Space Flight Center, Code 667, Greenbelt, MD 20771, USA \email{Aki.Roberge@nasa.gov}}
\altaffiltext{2}{Eureka Scientific, 2452 Delmer, Suite 100, Oakland, CA 96002, USA}
\altaffiltext{3}{Space Sciences Laboratory, University of California, 7 Gauss Way, Berkeley, CA 94720, USA}
\altaffiltext{4}{Kapteyn Astronomical Institute, University of Groningen, 9700 AV Groningen, Netherlands}
\altaffiltext{5}{Department of Terrestrial Magnitism, Carnegie Institution for Science, 5241 Broad Branch Road NW, Washington, DC 20015, USA}

%% Mark off your abstract in the ``abstract'' environment. In the manuscript
%% style, abstract will output a Received/Accepted line after the
%% title and affiliation information. No date will appear since the author
%% does not have this information. The dates will be filled in by the
%% editorial office after submission.

\begin{abstract}
We present \emph{Hubble Space Telescope} STIS far-UV spectra of the edge-on disk around 49~Ceti, one of the very few debris disks showing sub-mm CO emission.
Many atomic absorption lines are present in the spectra, most of which arise from circumstellar gas lying along the line-of-sight to the central star.
We determined the line-of-sight \ion{C}{1} column density, estimated the total carbon column density, and set limits on the \ion{O}{1} column density.
Surprisingly, no line-of-sight CO absorption was seen.
We discuss possible explanations for this non-detection, and present preliminary estimates of the carbon abundances in the line-of-sight gas.
The C/Fe ratio is much greater than the solar value, suggesting that \cet\ harbors a volatile-rich gas disk similar to that of $\beta$~Pictoris.
\end{abstract}


\section{Introduction}

Within about 10~Myr, gas-rich protoplanetary disks evolve into dusty, gas-poor debris disks composed of planetesimal destruction products.
The details of this evolution are poorly understood, including the primary gas dissipation mechanism and the changing composition of the disk material.
Disks may pass through a brief stage during which they contain both primordial material left over from star formation and secondary material from destruction of planetary bodies \citep[e.g.][]{Kospal:2013}.

The \object[HD9672]{49 Ceti} disk offers a rare opportunity to explore this possibility.
It is a nearby (61~pc) A1V star, recently identified as a co-moving member of the $\sim 40$~Myr-old Argus Association \citep{Zuckerman:2012} and long known to host a circumstellar (CS) dust disk \citep[e.g.][]{ Sadakane:1986}.
The modest amount of CS dust present led to its classification as a young debris disk.
However, it is one of only three debris disks known to show sub-mm CO emission \citep{Zuckerman:1995, Dent:2005, Hughes:2008}; the others are HD21997 \citep{Moor:2011} and $\beta$~Pictoris \citep{Dent:2014}.
Since abundant CO gas is characteristic of a primordial or transitional disk, this led to speculation that \cet\ might be a rare type of protoplanetary disk just at the end of its primordial gas dissipation phase \citep{Hughes:2008}.
However, this identification was difficult to reconcile with the age of the system and the detection of far-IR \ion{C}{2} emission from \cet\ but non-detection of \ion{O}{1} emission \citep{Roberge:2013}, which supported the suggestion that the CO gas is coming from icy comet-like bodies \citep{Zuckerman:2012}.

Here we report on far-ultraviolet (FUV) absorption spectroscopy of \cet. 
Since this disk is close to edge-on 
\citep[$i = 90^\circ \pm 5^\circ$;][]{Hughes:2008}, the line-of-sight to the central star intercepts disk material, as was shown by the detection of CS \ion{Ca}{2} gas in optical spectra of the star \citep{Montgomery:2012}.
The FUV is rich in atomic and molecular absorption lines, permitting a much more detailed inventory of the line-of-sight CS gas, as was done in the case of $\beta$~Pic \citep[e.g.][]{Roberge:2006}.
%
%In Section~\ref{sec:obs}, we describe the dataset and summarize its important %characteristics.
%Analyses of the atomic carbon and oxygen lines are presented in %Sections~\ref{sec:carbon} and \ref{sec:oxygen}.
%Surprisingly, no CO absorption lines were detected.
%An upper limit on the CO abundance is given in Section~\ref{sec:co} and compared %to the amount expected on the basis of the earlier sub-mm CO detection from %\citet{Hughes:2008}.
%Finally, in Section~\ref{sec:discussion}, we discuss possible explanations for %the detection of carbon and oxygen absorption but non-detection of CO, and give %preliminary estimates of the carbon abundances in the line-of-sight CS gas.

\section{Observations} \label{sec:obs} 

%% In a manner similar to \objectname authors can provide links to dataset
%% hosted at participating data centers via the \dataset{} command.  The
%% second curly bracket argument is printed in the text while the first
%% parentheses argument serves as the valid data set identifier.  Large
%% lists of data set are best provided in a table (see Table 3 for an example).
%% Valid data set identifiers should be obtained from the data center that
%% is currently hosting the data.
%%
%% Note that AASTeX interprets everything between the curly braces in the 
%% macro as regular text, so any special characters, e.g. "#" or "_," must be 
%% preceded by a backslash. Otherwise, you will get a LaTeX error when you 
%% compile your manuscript.  Special characters do not 
%% need to be escaped in the optional, square-bracket argument.
%%
%% \dataset[ADS/Sa.HST#Y0Q70101T]{HST FOS spectrum} that readers can access
%% via the links in the electronic edition.  

We observed \cet\ with the \emph{HST} Space Telescope Imaging Spectrograph (STIS) on \dataset[ads/Sa.HST#OBYE01010, ads/Sa.HST#OBYE01020, ads/Sa.HST#OBYE01030]{2013-08-11} and \dataset[ads/Sa.HST#OBYE02010, ads/Sa.HST#OBYE02020, ads/Sa.HST#OBYE02030]{2013-08-16}.
FUV spectra were acquired in two visits to search for changes in CS absorption lines, since the \ion{Ca}{2} absorption lines in \cet\ optical spectra showed signs of variability \citep{Montgomery:2012}.
The data were taken with E140H echelle grating and the $0\farcs2 \times 0\farcs09$ slit, providing spectra with $R = \lambda / \Delta \lambda = 228000$.
In each visit, the wavelength range from 1164~\AA\ to 1682~\AA\ was covered using three grating settings (i1271, c1416, and c1598), with exposure times of 1663~s, 540~s, and 1878~s.
The data were calibrated with the default CALSTIS v2.40 pipeline; the $1 \sigma$ relative velocity accuracy within one echelle order is 0.33~\kms\ for the i1271 spectra and 0.16~\kms\ for the c1416 and c1598 spectra\footnote{STIS Instrument Handbook, V13.0: \url{http://www.stsci.edu/hst/stis/documents/handbooks/currentIHB/cover.html} \label{foot:stis}}.
The absolute wavelength accuracy between echelle orders is 0.66~\kms\ for the i1271 spectra and 0.33~\kms\ for the c1416 and c1598 spectra.

We detected narrow absorption lines arising from the following species: \ci, \ion{C}{2}, \ion{C}{4}, \oi, \ion{Cl}{1}, \ion{S}{1}, \ion{Si}{2}, \ion{Al}{2}, and \ion{Fe}{2}.
Lines arising from excited fine-structure energy levels, which are not seen in spectra of the local interstellar (IS) medium, are present.
%These lines arise purely from CS gas with no significant IS contamination.
The wings of strong lines arising from abundant ionized species (\ion{C}{2} and \ion{C}{4}) showed significant variable absorption, indicating transient gas falling toward the central star at high velocity.
These features greatly resemble those seen in FUV spectra of $\beta$~Pic \citep[e.g.][]{Roberge:2000,Bouret:2002} and likely arise from the same phenomenon --
transits of star-grazing planetesimals \citep[e.g.][]{Beust:1990}.
Analysis of the variable features will be presented in a future paper.
No emission lines were seen, except Lyman~$\alpha$ emission very likely coming from terrestrial airglow.  
This shows that the central star is chromospherically inactive.

\section{Circumstellar Carbon}  \label{sec:carbon}

\subsection{Spin-forbidden \ion{C}{1} 1613.4~\AA\ line} \label{sub:1613}

\begin{figure}
\plotone{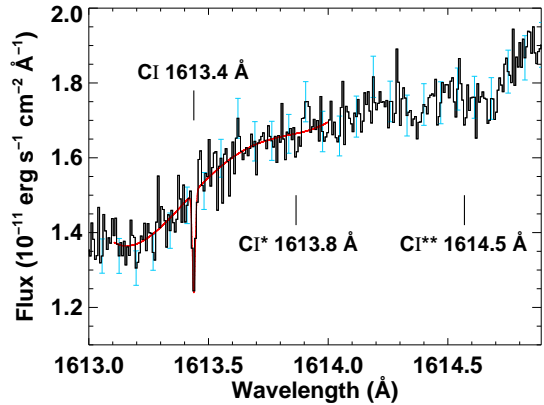}
\caption{The spin-forbidden \ion{C}{1} 1613~\AA\ triplet in the averaged \cet\ spectrum.  
Flux error bars are shown in light blue.
The red line shows the continuum model for the region around the 1613.4~\AA\ line, with the best-fitting Gaussian model for the absorption line superimposed. 
The two lines arising from excited fine-structure energy levels were not detected. \label{fig:1613} }
\end{figure}

The spin-forbidden \ci\ absorption line at 1613.4~\AA, arising from the ground energy level of \ci\ (\tpz), appears in the spectra from both visits.
This weak line remains unsaturated to high abundances, making it valuable for accurate measurements of large column densities \citep[e.g.][]{Roberge:2000}.
There is no hint of variability or multiple absorption components at different velocities. 
Therefore, the spectra from the two visits were averaged together to increase the $S/N$ (Figure~\ref{fig:1613}). 
To correct for any wavelength shift between the two exposures, the data were aligned before averaging by cross-correlating over a small wavelength region near the line. 
The 1613.4~\AA\ line is detected at the $5 \sigma$ level.
The two lines in the triplet arising from the 1st and 2nd excited fine structure energy levels (\ci$^*$ at 1613.8 and \ci$^{**}$ 1614.5~\AA) were not detected.

The continuum around the line was fit with a 4th-degree polynomial and the spectrum normalized.
We then measured the equivalent width (EQW) of the line and the column density in the ground energy level, $N(^3 \mathrm{P}_0) = (2.66 \pm 0.51) \times 10^{15} \ \mathrm{cm}^{-2}$.
This is about 44\% of the line-of-sight \ci\ \tpz\ column density towards \bp\ \citep{Roberge:2000}.
The line was fit with a Gaussian to determine the heliocentric velocity of the gas and the Doppler broadening parameter, $b = 1.32 \pm 0.35$~\kms; the error is the statistical uncertainty added in quadrature with the uncertainty in the relative calibration of the wavelength scale\footnotemark[\ref{foot:stis}]. 

The velocity is $v_{\mathrm{CI}} = 11.59 \pm 0.40$~\kms, where the error includes the uncertainty in the absolute wavelength calibration\footnotemark[\ref{foot:stis}].
The velocity is close to the radial velocity of the central star \citep[$v_\star = 12.2$~\kms;][]{Hughes:2008}.
Unfortunately, the star's velocity is also close to the velocity of the local IS cloud in the direction of \cet\ \citep[$v_\mathrm{LIC} = 11.0 \pm 1.3$~\kms;][]{Redfield:2008}.
This makes it impossible to identify the \ci\ absorption as being purely CS on the basis of velocity alone. 
However, the favored ionization state for carbon in the diffuse ISM is \ion{C}{2} and local IS \ion{C}{1} abundances are very low \citep[$N(^3 \mathrm{P}_0) \lesssim 10^{12}$~\cm;][]{Welsh:2010}.
Therefore, this large amount of \ion{C}{1} gas towards \cet\ is likely to be almost entirely CS.

\subsection{\ion{C}{1} 1560~\AA\ multiplet \label{sub:1560}}

\begin{figure}
\plotone{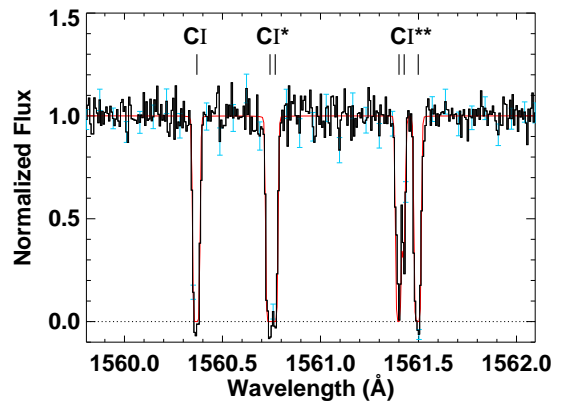}
\caption{The strong dipole-allowed \ion{C}{1} 1560~\AA\ multiplet in the averaged, normalized \cet\ spectrum.
%The normalized spectra are plotted with black lines; flux error bars are shown in light %blue. 
The red line shows the best-fitting absorption model. \label{fig:1560} }
\end{figure}

Several strong dipole-allowed multiplets of \ion{C}{1} are visible in the spectra; the  multiplet that lies in a region of the spectrum with the highest continuum $S/N$ is at 1560~\AA\ (Figure~\ref{fig:1560}).
Lines arising from the first three fine structure energy levels are clearly detected, so we are able to constrain the excitation temperature and the total column density.
The \ci\ 1560~\AA\ multiplet did not vary significantly between the two visits, and there are no obvious signs of multiple velocity components.
Therefore, the spectra from the two visits were aligned in wavelength, then averaged together to increase the $S/N$.

We created models of the multiplet containing a single velocity component, assuming LTE for the population of the fine structure levels, and using Voigt line profiles appropriate for saturated lines. 
The free parameters in the models are the total column density in the $^3 \mathrm{P}$ ground term ($N(^3 \mathrm{P})$), the excitation temperature ($T_\mathrm{ex}$), $b$, and $v$.
The best parameters were found by $\chi^2$ minimization, varying all free parameters over large ranges of parameter space. 
The uncertainties on the best parameters were determined by projecting the $\chi^2$ contours, which were well-behaved with only a single minimum seen.
The total column density is $N(^3 \mathrm{P}) = 4.47^{+1.15}_{-0.43} 
\times 10^{15}$~\cm, which is within the upper limit on the total column density determined from the non-detections of the excited lines in the 1613~\AA\ triplet.
The excitation temperature is $T_\mathrm{ex} = 38.0^{+8.0}_{-3.8}$~K.
The Doppler broadening parameter is $b = 1.65 \pm 0.17$~\kms\ and the velocity is 
$v = 11.00 \pm 0.33$~\kms, which agrees with the velocity determined from the 1613.4~\AA\ line within the errors.

\subsection{Total carbon} \label{sub:carbon}

A strong saturated doublet arising from \ion{C}{2} at 1334~\AA\ is seen in the data.
Ideally, measurement of the CS \ion{C}{2} column density would allow us to calculate the total abundance of CS carbon along the line-of-sight to \cet.
Unfortunately, the line arising from the ground energy level will contain blended IS contamination.
Furthermore, both \ion{C}{2} features varied between the two visits, showing that some of the gas likely arises from star-grazing planetesimals. 
Modeling of these lines will be very difficult and is deferred to a later paper.

To roughly estimate the total CS carbon column density, we make use of elemental neutral fractions calculated for a similar environment, the disk around the A5V star \bp.
Ionization balance calculations indicated that the \bp\ carbon gas is about 60\% neutral \citep{Fernandez:2006}, which was confirmed by observation \citep{Roberge:2006}.
The \bp\ and \cet\ central stars are both A-type and their disks have a similar total dust abundance (based on their fractional dust luminosities).
While sub-mm observations suggest that the \cet\ disk as a whole contains more CO than \bp\ \citep{Hughes:2008,Dent:2014}, the two CS environments appear otherwise similar.
\citet{Fernandez:2006} also provided calculated neutral fractions for \bp-like gas disks with different central stars.
For an A0V star (close to the A1V spectral type of \cet), the carbon neutral fraction should be about 1\%, giving a total CS carbon column density along the line-of-sight of $\sim 4.5 \times 10^{17}$~\cm.

\section{Circumstellar Oxygen} \label{sec:oxygen}

\begin{figure}
\plotone{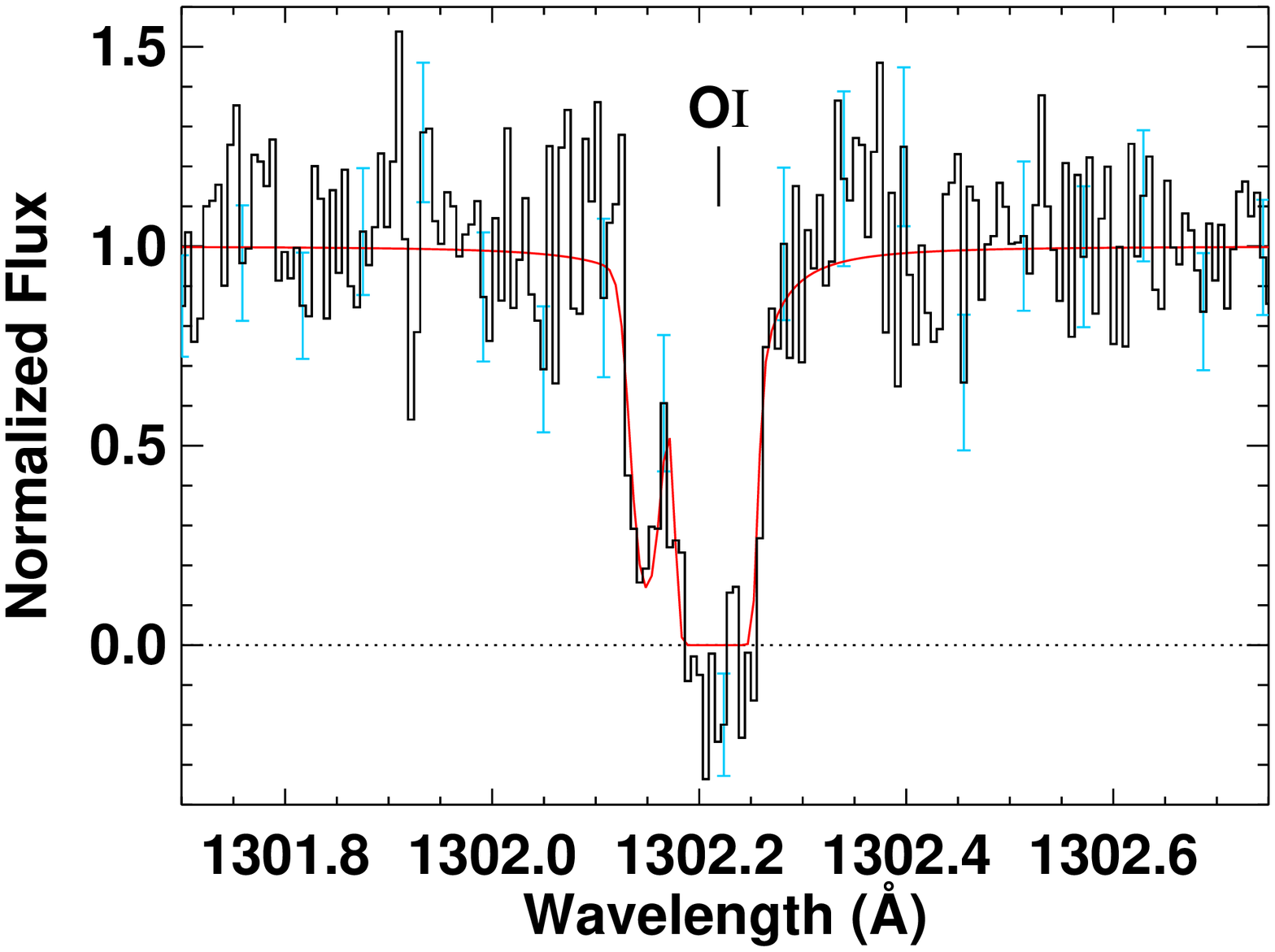}
\plotone{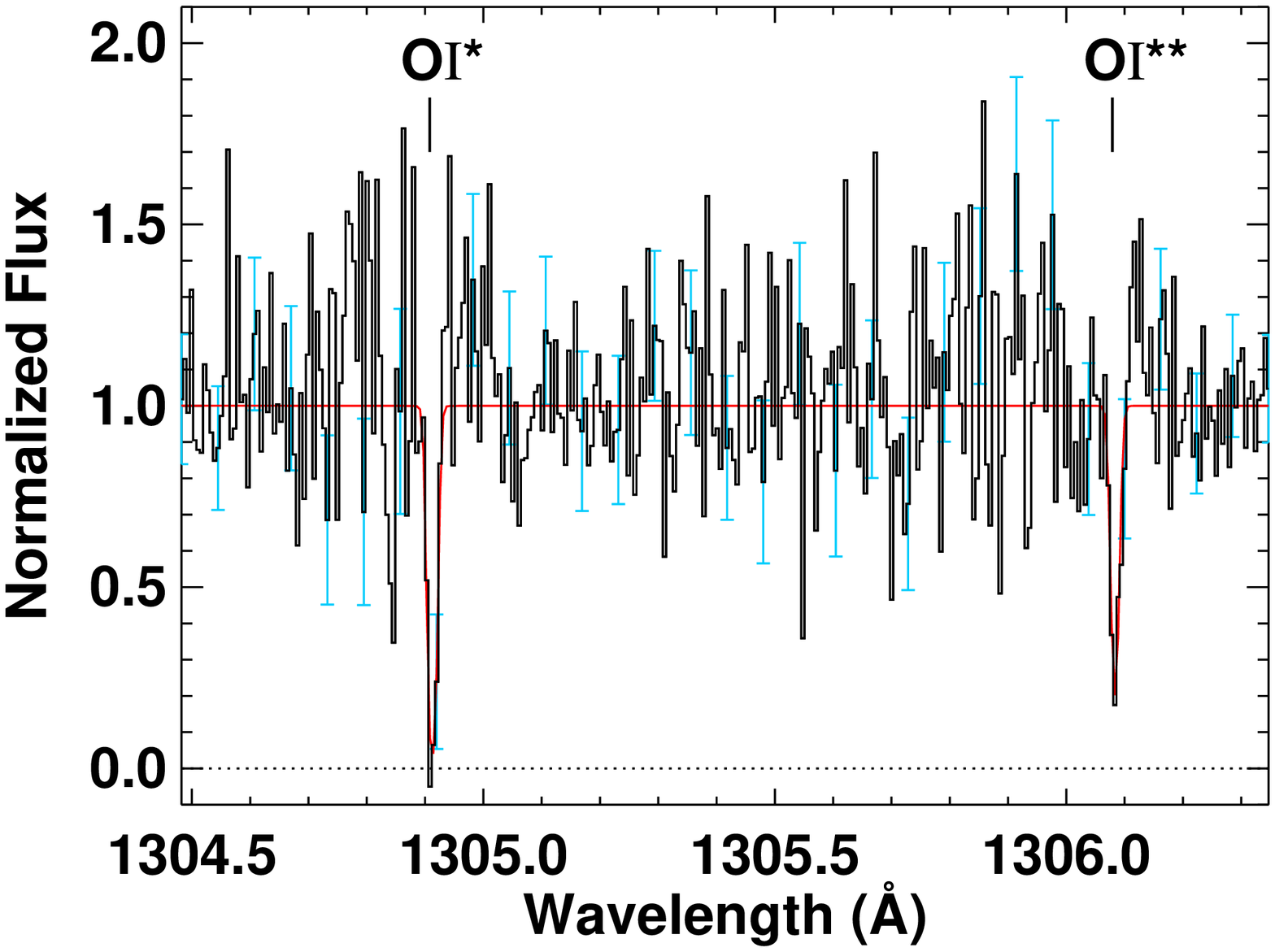}
\caption{The \ion{O}{1} 1302~\AA\ multiplet in the averaged, normalized \cet\ spectrum.
%The normalized spectra are plotted with black lines; flux error bars are shown in light %blue. 
The top panel shows the saturated, blended \oi\ line at 1302.2~\AA, while the bottom panel shows the weaker \ois\ and \oiss\ lines arising purely from CS gas. 
The red lines show the best-fitting absorption models. \label{fig:oi} }
\end{figure}

Three lines of \oi\ at 1302.2, 1304.9, and 1306.0~\AA\ were seen in the spectra (Figure~\ref{fig:oi}), arising from the three fine structure energy levels of the ground term (\tpt, \tpo, and \tpz).
The line arising from the ground level is very likely contaminated with blended IS absorption; furthermore, there appears to be an additional blueshifted component in the absorption.
While there was no \emph{significant} variation in the blueshifted feature between the two visits, there is a hint that it changed in depth, indicating that it is likely due to a star-grazing comet. 
The complexity, strength, and relatively low $S/N$ of the \oi\ 1302.2~\AA\ absorption feature make it challenging to model.

However, the \ois\ and \oiss\ lines arising from the excited levels come solely from CS gas, since the column density of \ois\ in the local ISM is $\lesssim 2 \times 10^{12}$~\cm\ \citep{Welsh:2010}.
These lines did not vary significantly between the two visits and there are no obvious signs of multiple velocity components.
We modeled these lines first, to illuminate the velocity structure of the CS \oi\ component and then use that information to aid modeling of the saturated, blended \oi\ 1302.2~\AA\ line.

For the \ois\ and \oiss\ lines, we performed a $\chi^2$ minimization analysis similar to the one described for the \ci\ 1560~\AA\ multiplet.
Models for the two lines were created with a single velocity component. 
One difference is that we did not assume LTE for these models.
Unlike \ci, the separations between the fine structure energy levels of \oi\ are large and very high temperature and density are required for \oi\ gas to be in LTE.
The free parameters in the models are $N(^3 \mathrm{P}_1)$, $N(^3 \mathrm{P}_0)$, $b$, and $v$. 
Again, the contours of $\chi^2$ were well-behaved with only a single minimum seen.
We found $N(^3 \mathrm{P}_1) = (7.08 \pm 0.92) \times 10^{13}$~\cm, 
$N(^3 \mathrm{P}_0) = (3.16 \pm 0.91) \times 10^{13}$~\cm, 
$b = 1.75 \pm 0.34$~\kms, and $v = 12.6 \pm 0.70$~\kms.
The velocity of the gas agrees very well with the velocity of \cet\ (12.2~\kms), consistent with a gas disk in Keplerian rotation about the star.

We then created models of the \oi\ 1302.2~\AA\ line with three velocity components:
a CS component with the $b$ and $v$ values determined from the \ois\ and \oiss\ lines, an IS component at the velocity of the local IS cloud toward \cet\ \citep[11~\kms;][]{Redfield:2008} and the mean $b$ value for \oi\ in the local ISM \citep[3.24~\kms;][]{Redfield:2004}, and a blueshifted component.
The free parameters in the models were $N_\mathrm{CS}(^3 \mathrm{P}_2)$, $N_\mathrm{IS}(^3 \mathrm{P}_2)$, $N_\mathrm{BS}(^3 \mathrm{P}_2)$, $b_\mathrm{BS}$, and $v_\mathrm{BS}$. 
As before, $\chi^2$ minimization was performed over large ranges of parameter space; the best-fitting model appears in Figure~\ref{fig:oi}.
However, the $\chi^2$ contours were not well-behaved this time, with signs of multiple minima and open-ended contours for $N_\mathrm{CS}(^3 \mathrm{P}_2)$. 
This indicates that the modeling is too degenerate to draw strong conclusions about the best parameters.

However, we were able to set a $3 \sigma$ lower limit on the \emph{combined} IS/CS column density, $N_\mathrm{IS/CS}(^3 \mathrm{P}_2) \gtrsim 2 \times 10^{16}$~\cm. 
We could also set a $3 \sigma$ upper limit on $N_\mathrm{CS}(^3\mathrm{P}_2) \lesssim 1 \times 10^{17}$~\cm.
The upper limit on $N_\mathrm{CS}(^3\mathrm{P}_2)$ that we calculated from non-detection of a weak spin-forbidden \oi\ line at 1355.6~\AA\ is similar ($\leq 1.85 \times 10^{17}$~\cm).

\pagebreak
\section{Upper Limit on Carbon Monoxide}  \label{sec:co}

\begin{figure}
\plotone{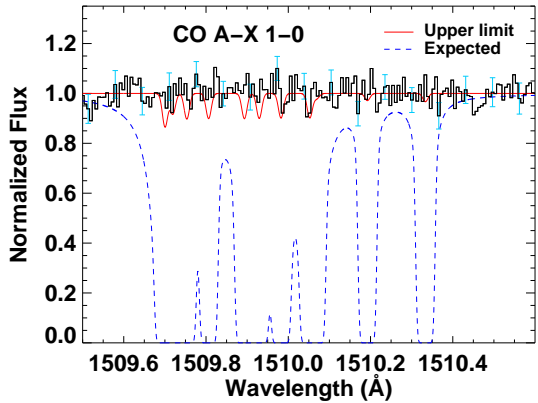}
\caption{Non-detection of line-of-sight CO absorption.  
The black line shows the averaged, normalized spectrum in the region of the $A-X$ (1-0) band.
The red line is a model for the absorption band with the $3 \sigma$ upper limit on the column density in each rotational level.
The blue dashed line is a model for the expected absorption based on analysis of the observed sub-mm CO emission from \cet\ \citep{Hughes:2008}.
\label{fig:co} }
\end{figure}

No absorption was seen in any of the bands of the CO fourth positive system ($A \ ^1 \Pi - X \ ^1 \Sigma^+$).
Figure~\ref{fig:co} shows the region around the $A-X$ (1-0) band.
To calculate upper limits on CO in the five lowest rotational levels ($J_l =0 - 4$), we fit the continua in the regions of the (0-0), (1-0), (2-0), and (3-0) bands with polynomials, then normalized the spectra.
The EQW and error was calculated for every line arising from the $J_l = 0, 1, 2, 3, \ \mathrm{and} \ 4$ levels, then converted into a column density and uncertainty.
For each rotational level, the line with the smallest EQW uncertainty was used to set the $3 \sigma$ upper limit on the column density (Table~\ref{tab:co}).

The non-detection of CO absorption in these spectra was surprising, given the edge-on inclination of the \cet\ CO disk observed in emission with the Submillimeter Array \citep[$i = 90^\circ \pm 5^\circ$;][]{Hughes:2008}.
Based on models of the observed sub-mm CO emission in \citet{Hughes:2008}, we expected a CO excitation temperature of 250~K and a line-of-sight column density of $2.8 \times 10^{18}$~\cm, which is about 5 orders of magnitude above our $3 \sigma$ upper limit on the total line-of-sight absorption.
A model of the expected absorption in the CO (1-0) band appears in Figure~\ref{fig:co}.

\section{Discussion} \label{sec:discussion}

\subsection{Why is C absorption seen, but no CO?}

First, for there to be bright sub-mm CO emission from \cet\ but no line-of-sight CO absorption, the molecular gas disk must not be as edge-on as was thought.
This will be confirmed or denied by a new high spatial resolution CO map of \cet\ obtained with ALMA (Hughes et al., in preparation).
Once the true inclination of the \cet\ molecular gas disk is established, the lack of line-of-sight CO will provide a strong constraint on the scale height of the gas.

Second, the detection of C and O absorption along the line-of-sight to the central star but non-detection of CO absorption shows that the spatial distributions of atomic and molecular gas in this disk are different.
The atomic absorption lines contain a component close to the velocity of the central star, which does not significantly vary between the two days of observation. 
This component greatly resembles the so-called ``stable'' component seen in  spectra of \bp, which is associated with a gas disk in Keplerian rotation about the star \citep[e.g.][]{Lagrange:1998}.
It appears likely that \cet\ too has a stable component.

Such a stable component could arise in a largely atomic gas disk rotating about the star, as with \bp. 
This disk would have to have a larger scale height than the molecular disk, which could happen if much of the atomic gas is produced from comets with a wide range of inclinations.
But given the presence of abundant CO, one must also consider the possibility that the atomic gas is a dissociated and ionized skin over the molecular gas disk.
% 
%Such layered structure is predicted and observed for protoplanetary disks, which are %gas-rich and optically thick.
%However, the amount of dust present around \cet\ is too modest to make the disk optically %thick to dissociating radiation.
%Whether the CO can shield itself from rapid dissociation is currently unknown, due to %insufficient knowledge of the exact gas spatial distribution.
%
Hopefully, ALMA maps of CO emission (recently acquired) and \ion{C}{1} fine structure emission (upcoming in Cycle~2) will reveal the spatial distributions of molecular and atomic gas in the disk.

\begin{table}
\begin{center}
\caption{$3 \sigma $ upper limits on line-of-sight CO absorption \label{tab:co}}
\begin{tabular}{lc}
\hline \hline
Lower & Column density \\
level & (\cm) \\
\hline
$J = 0$ & $\leq 2.9 \times 10^{12}$ \\
$J = 1$ & $\leq 5.8 \times 10^{12}$ \\
$J = 2$ & $\leq 5.6 \times 10^{12}$ \\
$J = 3$ & $\leq 5.4 \times 10^{12}$ \\
$J = 4$ & $\leq 5.2 \times 10^{12}$ \\
\hline
Total & $\leq 2.5 \times 10^{13}$ \\
\hline
\end{tabular}
\end{center}
\end{table}

\subsection{Preliminary carbon abundances in the CS gas}

Given its high first ionization energy, the bulk of the oxygen should be in the neutral state.
Therefore, the upper limit on the total line-of-sight CS oxygen column density is $N_\mathrm{O} \lesssim 10^{17}$~\cm.
The approximate C/O number ratio is then $\gtrsim 4.5$, which is about 9 times greater than the solar C/O ratio \citep[0.5;][]{Lodders:2003}.
This value should be taken with a grain of salt, however, since we only estimated the CS \ion{C}{2} column density.
It needs to be confirmed, if possible, by modeling the complex, saturated, and variable \ion{C}{2} absorption features in our spectra.

\citet{Malamut:2014} analyzed two strong near-UV \ion{Fe}{2} absorption lines towards \cet\ and found two velocity components, one IS and one CS. 
The CS component had $v = 13.65 \pm 0.15$~\kms, $b = 2.50 \pm 0.28$~\kms, and $N = 1.86^{+0.48}_{-0.38} \times 10^{13}$~\cm.
The neutral fraction for \ion{Fe}{2} around \cet\ should be $3 \times 10^{-6}$, assuming a \bp-like gas disk around an A0V star \citep{Fernandez:2006}.
This is supported by non-detection of an \ion{Fe}{1} line at 3859.9~\AA\ in optical spectra of \cet\ (Welsh \& Montgomery, in preparation).
So nearly all the Fe should be in the 1st ionized state and the total column density of CS iron is $1.86 \times 10^{13}$~\cm.

Therefore, the C/Fe number ratio in the line-of-sight CS gas is $\sim  24000$.
The solar C/Fe number ratio is 8.7 \citep{Lodders:2003}, so the CS carbon along the line-of-sight to \cet\ appears to be overabundant by a factor of about 2800.
Even if we ignore \ion{C}{2} and take the reliable \ion{C}{1} column density to be the total carbon column density, the C/Fe ratio is still $\sim 28$ times the solar value.

An extreme carbon overabundance relative to iron, together with a more normal C/O ratio, is also seen in the \bp\ CS gas \citep{Roberge:2006,Xie:2013}.
Both gas disks appear to be volatile-rich.
For a long time, it was not understood why radiation pressure does not rapidly blow away much of the \bp\ CS gas \citep{Lagrange:1998}.
\citet{Fernandez:2006} showed that the moderately ionized \bp\ gas couples into a single ionic fluid with an effective radiation pressure coefficient, and that a carbon overabundance lowers the coefficient, keeping the whole gas disk bound to the star.
The carbon overabundance now seen in \cet\ may play a similar role.

In this scenario, small velocity shifts between different atomic species are expected, as neutral species may be slightly accelerated by radiation pressure before being ionized and joining the ionic fluid.
This may explain why the CS \ci\ velocity we measured is slightly blueshifted with respect to the CS \ion{Fe}{2} velocity measured by \citet{Malamut:2014}.
Detailed calculations of the \cet\ gas dynamics and comparison to the exact radial velocities of the observed species will provide a strong test of this theory.

\acknowledgments

Support for program number GO-12901 was provided by NASA through a grant from the Space Telescope Science Institute, which is operated by the Association of Universities for Research in Astronomy, Inc., under NASA contract NAS5-26555.
A.~R.\ also acknowledges support by the Goddard Center for Astrobiology, part of the NASA Astrobiology Institute.

%% To help institutions obtain information on the effectiveness of their
%% telescopes, the AAS Journals has created a group of keywords for telescope
%% facilities. A common set of keywords will make these types of searches
%% significantly easier and more accurate. In addition, they will also be
%% useful in linking papers together which utilize the same telescopes
%% within the framework of the National Virtual Observatory.
%% See the AASTeX Web site at http://aastex.aas.org/
%% for information on obtaining the facility keywords.

%% After the acknowledgments section, use the following syntax and the
%% \facility{} macro to list the keywords of facilities used in the research
%% for the paper.  Each keyword will be checked against the master list during
%% copy editing.  Individual instruments or configurations can be provided 
%% in parentheses, after the keyword, but they will not be verified.

{\it Facilities:} \facility{HST (STIS)}.

%% Appendix material should be preceded with a single \appendix command.
%% There should be a \section command for each appendix. Mark appendix
%% subsections with the same markup you use in the main body of the paper.

%% Each Appendix (indicated with \section) will be lettered A, B, C, etc.
%% The equation counter will reset when it encounters the \appendix
%% command and will number appendix equations (A1), (A2), etc.

%% \appendix

%\bibliographystyle{apj}
%\bibliography{akir}

\end{document}